\documentclass[aps,prl,amsmath,showpacs,preprintnumbers,twocolumn,amssymb,amsmath,superscriptaddress]{revtex4-1}

\pdfoutput=1
\usepackage{graphicx}
\usepackage{mathptmx, textcomp}
\usepackage[latin1]{inputenc}
\usepackage{tabularx}
\usepackage{units}
\usepackage{color}

\begin{document}
\title{Extended Bose-Hubbard Models with Ultracold Magnetic Atoms}

\author{S. Baier}
\affiliation{Institut f\"ur Experimentalphysik, Universit\"at Innsbruck, Technikerstra{\ss}e 25, 6020 Innsbruck, Austria}
\author{M. J. Mark}
\affiliation{Institut f\"ur Experimentalphysik, Universit\"at Innsbruck, Technikerstra{\ss}e 25, 6020 Innsbruck, Austria}
\affiliation{Institut f\"ur Quantenoptik und Quanteninformation, \"Osterreichische Akademie der Wissenschaften, 6020 Innsbruck, Austria}
\author{D. Petter}
\affiliation{Institut f\"ur Experimentalphysik, Universit\"at Innsbruck, Technikerstra{\ss}e 25, 6020 Innsbruck, Austria}
\author{K. Aikawa}
\altaffiliation{Current adress: Department of Physics, Graduate School of Science and Engineering, Tokyo Institute of Technology, Meguro-ku, Tokyo, 152-8550 Japan}
\affiliation{Institut f\"ur Experimentalphysik, Universit\"at Innsbruck, Technikerstra{\ss}e 25, 6020 Innsbruck, Austria}
\author{L. Chomaz}
\affiliation{Institut f\"ur Experimentalphysik, Universit\"at Innsbruck, Technikerstra{\ss}e 25, 6020 Innsbruck, Austria}
\affiliation{Institut f\"ur Quantenoptik und Quanteninformation, \"Osterreichische Akademie der Wissenschaften, 6020 Innsbruck, Austria}
\author{Z. Cai}
\affiliation{Institut f\"ur Quantenoptik und Quanteninformation, \"Osterreichische Akademie der Wissenschaften, 6020 Innsbruck, Austria}
\author{M. Baranov}
\affiliation{Institut f\"ur Quantenoptik und Quanteninformation, \"Osterreichische Akademie der Wissenschaften, 6020 Innsbruck, Austria}
\author{P. Zoller}
\affiliation{Institut f\"ur Quantenoptik und Quanteninformation, \"Osterreichische Akademie der Wissenschaften, 6020 Innsbruck, Austria}
\affiliation{Institut f\"ur Theoretische Physik, Universit\"at Innsbruck, Technikerstra{\ss}e 21A, 6020 Innsbruck, Austria}
\author{F. Ferlaino}
\affiliation{Institut f\"ur Experimentalphysik, Universit\"at Innsbruck, Technikerstra{\ss}e 25, 6020 Innsbruck, Austria}
\affiliation{Institut f\"ur Quantenoptik und Quanteninformation, \"Osterreichische Akademie der Wissenschaften, 6020 Innsbruck, Austria}
\date{\today}


\begin{abstract}
The Hubbard model underlies our understanding of strongly correlated materials. While its standard form only comprises interaction between particles at the same lattice site, its extension to encompass long-range interaction, which activates terms acting between different sites, is predicted to profoundly alter the quantum behavior of the system. We realize the extended Bose-Hubbard model for an ultracold gas of strongly magnetic erbium atoms in a three-dimensional optical lattice. Controlling the orientation of the atomic dipoles, we reveal the anisotropic character of the onsite interaction and hopping dynamics, and their influence on the superfluid-to-Mott insulator quantum phase transition. Moreover, we observe nearest-neighbor interaction, which is a genuine consequence of the long-range nature of dipolar interactions. Our results lay the groundwork for future studies of novel exotic many-body quantum phases.
\end{abstract}

\maketitle


Dipolar interactions, reflecting the forces between a pair of magnetic
or electric dipoles, account for many physically and biologically
significant phenomena. These range from novel phases appearing at
low temperatures in quantum many-body systems\,\cite{Lahaye2009,Baranov2012},
liquid crystals and ferrofluids in soft condensed matter physics\,\cite{Gennes1995,Rosensweig1985},
to the mechanism underlying protein folding\,\cite{Dill1990}. The
distinguishing feature of dipole-dipole interactions (DDI) is their
long-range and anisotropic character \cite{note3}: a pair of dipoles oriented in
parallel will repel each other, while the interaction between two
head to tail dipoles will be attractive. While remarkable progress
has been made with gases of polar molecules\,\cite{Yan2013} and
Rydberg ensembles\,\cite{Schauss2015} comprising electric dipoles,
it is the recent experimental advances in creating quantum degenerate
gases of bosonic and fermionic magnetic atoms, including Cr\,\cite{Griesmaier2005,dePaz2013,Naylor2015}
and the Lanthanides Er\,\cite{Aikawa2012bec} and Dy\,\cite{Lu2011},
which have now opened the door to a study of magnetic dipolar interactions,
and their unique role in Hubbard dynamics of a quantum lattice gas.

Ultracold Lanthanide atoms with their open electronic f-shells, and
their anisotropic interactions are characterized by unconventional
low energy scattering properties, including the proliferation of Feshbach
resonances\,\cite{Frisch2014qci}. This complexity of Lanthanides
manifests itself in quantum many-body dynamics: by preparing quantum
degenerate Lanthanide gases in optical lattices we realize \emph{extended
	Hubbard models} for bosonic and fermionic atoms. Here, in addition
to the familiar single particle tunneling and isotropic onsite interactions
(as for contact interactions in Alkali) dipolar interactions give
rise to anisotropic onsite and nearest-neighbor (offsite) interactions
(NNI), and density-assisted tunneling (DAT) \cite{note6}. Such extended Hubbard
models have been studied extensively in theoretical condensed matter
physics and quantum material science\,\cite{Micnas1990,Zaanen1999},
and it is the competition between these unconventional Hubbard interactions,
which underlies the prediction of exotic quantum phases such as supersolids,
stripe and checkerboard phases\,\cite{Scalettar1995,Scarola2005,Wessel2005,Yi2007,Pollet2010,Capogrosso2010}.

Here we report a first observation of the unique manifestations of
magnetic dipolar interactions in extended Hubbard dynamics. These
observations are enabled by preparing an ultracold sample of bosonic
Er atoms in an three-dimensional (3D) optical lattice. It is the control
of the optical lattice via laser parameters in combination with a
flexible alignment of the magnetic dipoles in an external magnetic
field, which allows us to reveal and explore the anisotropic onsite
and offsite interactions. Measurements of the excitation spectrum
in the Mott insulator state, and of the superfluid-to-Mott-insulator
(SF-to-MI) quantum phase transition are employed as a tool to detect
these interactions and their competitions.

\begin{figure*}[t]
	\includegraphics[width=1.5\columnwidth] {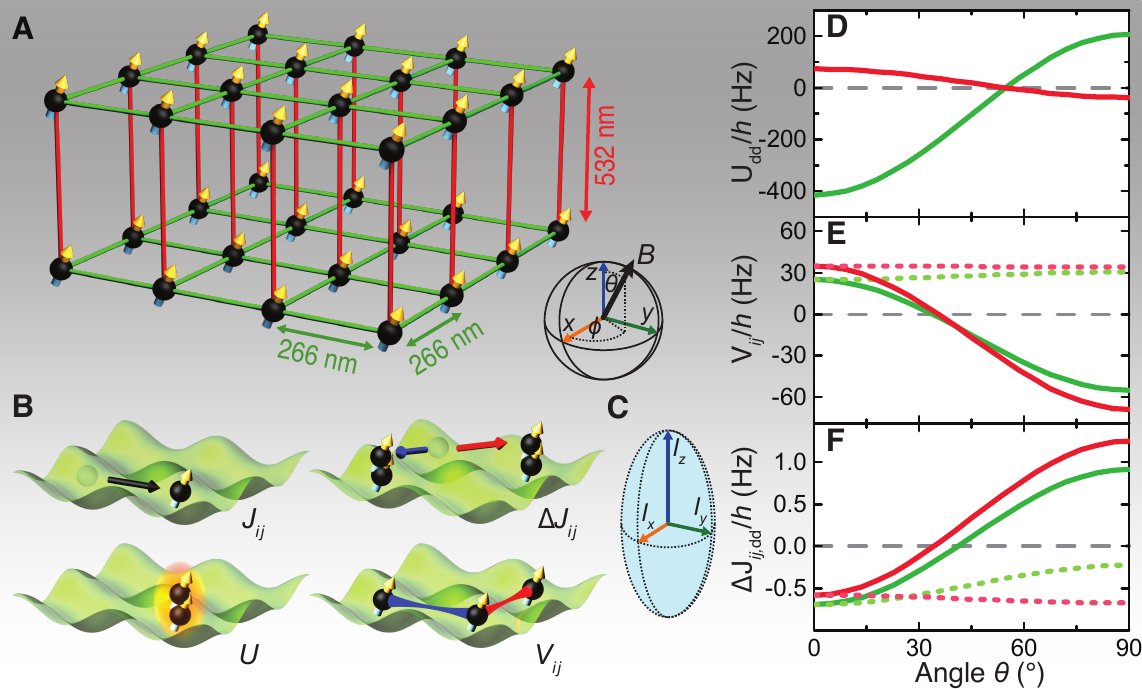}
	\caption{(color online) Magnetic dipoles in a 3D optical lattice. (A) Schematic of our lattice geometry, where the lattice constants are indicated. The dipole orientation, given by the polarizing magnetic field $B$, is quantified by the polar angles $\theta$ and $\phi$ with respect to our coordinate system. (B) Illustration of the contributing terms in the eBH model: Tunneling matrix element $J_{ij}$, DAT matrix elements $\Delta J_{ij}$, onsite interaction $U$,  and NNI $V_{ij}$. (C) Illustration of the definition of the onsite aspect ratio. (D to F) Calculated values of the DDI-dependent terms as a function of $\theta$ for $\phi=0^\circ$  and typical experimental parameters $(s_{x},s_{y},s_{z})\,=\,(15,15,s_{z})$ with $s_{z}$ set by the $\rm{AR}$ for the cases $\rm{AR}\,=\,1$ (red) and $\rm{AR}\,=\,2$ (green). (C) shows $U_{\mathrm{dd}}$. (D) gives $V_{ij=x}$ (solid lines) and $V_{ij=y}$ (dotted lines), the NNI for bond direction $x$ and $y$ respectively. (E) shows $\Delta J_{x,{\rm dd}}$ (solid lines) and $\Delta J_{y,{\rm dd}}$ (dotted lines), the values of $\Delta J_{ij,{\rm dd}}$ for hopping direction $x$ and $y$ respectively. The dashed lines indicate the case without DDI. $U_{\rm s}$ and $J_{ij}$ are independent on $\theta$ and their values for the two configurations considered are $U_{\rm s}\,=\,3749\,\rm Hz$ ($1775\,\rm Hz$) for ${\rm AR}\,=\,1$ (2) and $J_{ij}\,=\,27\,\rm Hz$.}
\label{fig:fig_1}
\end{figure*}

In our experiment an ultracold dipolar gas of $^{168}$Er atoms is
prepared in a 3D optical lattice. The atoms are spin-polarized in
their lowest Zeeman sublevel\,\cite{Aikawa2012bec} and feature a
magnetic moment $\mu$ of $7$ Bohr magneton. The experiment starts by adiabatically
loading a Bose-Einstein condensate (BEC) of about $1.5\times10^{5}$
atoms from an optical dipole trap (ODT) into the optical lattice.
The lattice is created by two retroreflected $532$-$\mathrm{nm}$
laser beams, defining the horizontal $xy$-plane, and one $1064$-$\mathrm{nm}$
beam, nearly collinear with the vertical ($z$) direction given by
gravity (Fig.\,1A) (Supplementary Materials). The lattice has a cuboid unit cell with lattice constants
$d_{x,y}\,=\,266\,{\rm nm}$ and $d_{z}\,=\,532\,{\rm nm}$, which correspond for Er to the recoil energies $\mathrm{E}_{\mathrm{R},x}=\,\mathrm{E}_{\mathrm{R},y}=h\times\,4.2\,{\rm kHz}$
and $\mathrm{E}_{\mathrm{R},z}\,=h\times\,1.05\,{\rm kHz}$, 
$h$ being Planck's constant. In addition, the lattice can be controlled by independently changing the depths  associated with the lattice beams in each direction, $(s_{x},s_{y},s_{z})$, measured here in units of the corresponding recoil energies.
The dipole orientation, quantified by the
polar angles $\theta$ and $\phi$ (inset Fig.\,1A), is varied by changing
the direction of the polarizing magnetic field (Supplementary Materials). By changing
the lattice depths we can prepare the Er atoms in a Mott insulator
state, driving a SF-to-MI phase transition, as described below.


The dynamics of Er atoms in the optical lattice is described by an
extended Bose-Hubbard (eBH) model with Hamiltonian\,\cite{Mazzarella2006,Dutta2015}
\begin{equation}
\begin{split}
H= & -\sum_{\langle ij\rangle}\left[\left(J_{ij}+\Delta J_{ij}(n_{i}+n_{j}-1)\right)b_{i}^{\dag}b_{j}^{\phantom{\dagger}}+{\rm h.c.}\right] \\
& +\frac{U}{2}\sum_{i}n_{i}(n_{i}-1)+\sum_{ij}V_{ij}n_{i}n_{j}.\label{eBHM}
\end{split}
\end{equation}
Here $b_{i}^{\dag}$ ($b_{i}^{\phantom{\dagger}}$) are the bosonic
creation (annihilation) operators of atoms at site $i$, $n_{i}=b_{i}^{\dag}b_{i}^{\phantom{\dag}}$
is the associated number operator, and $\langle ij\rangle$ denotes
pairs of adjacent sites. The first term in Eq.\,1 includes the single-particle hopping, with amplitudes $J_{ij}$ reflecting the anisotropy of the optical lattice. Interactions manifest themselves in an onsite
interactions $U$, offsite interactions $V_{ij}$ (approximated as nearest-neighbor-interaction (NNI)),
and a density-assisted-tunneling term (DAT) $\Delta J_{ij}$\,\cite{Luehmann2012,Juergensen2014ood}. 
All terms of the eBH are illustrated in Fig.\,1B. As discussed in the Supplementary Materials (see also \cite{Dutta2015}), the onsite interaction
$U$ and DAT $\Delta J_{ij}$ have contributions from both the short-range
part of the interatomic interaction ($U_{\mathrm{s}}$ and $\Delta J_{ij,{\rm s}}$),
which is proportional to the $s$-wave scattering length $a_{\mathrm{s}}$,
and from the long-range DDI ($U_{\mathrm{dd}}$ and $\Delta J_{ij,{\rm dd}}$), which is proportional to $\mu^2$.
On the other hand, $V_{ij}$ originates entirely from the long-range DDI. This mechanism for NNI 
is in marked contrast to, for example, Heisenberg spin-spin interaction between atoms at neighboring sites $i,j$, which arises from superexchange processes in Hubbard dynamics in second order virtual hopping processes $\sim J_{ij}^{2}/U$ in the limit of large onsite interaction\,\cite{Trotzky2008,Auerbach2012}.

The unique and characteristic feature of our many-body system is
contained in the angular dependence of $U$, $\Delta J_{ij}$, and
$V_{ij}$, reflecting the DDI in our eBH model. It reveals itself
prominently in combination with an {\em anisotropic Wannier function} at
a given lattice site reflecting the local density distribution (Fig.\,1C). The aspect ratio ${\rm AR}$ of the Wannier function can be
changed by imposing unequal lattice depths $(s_{x},s_{y},s_{z})$. In our experiment $s_x=s_y$, such that $z$ is the anisotropy axis. We define ${\rm AR} =l_z/l_{x,y}$, where $l_z$ ($l_x=l_y$) is the harmonic oscillator length along the $z$ ($xy$) direction of the local atomic well\,\cite{note2}. The relative weight between the attractive and repulsive
contribution to $U_{\mathrm{dd}}$ can be tuned by changing the dipole
orientation relative to the anisotropy axis of the onsite density
distribution, and the $\mathrm{AR}$ (Fig.\,1D).  In contrast, the NNI $V_{ij}$ is controlled
through the orientation of the dipoles with respect to the bond direction
$ij$ (Fig.\,1E). Finally, $\Delta J_{ij, {\rm dd}}$ depends both on the orientation
of the dipoles relative to the bond and anisotropy axes, and on
the ${\rm AR}$ (Fig.\,1F).


\begin{figure*}[t]
	\includegraphics[width=1.5\columnwidth] {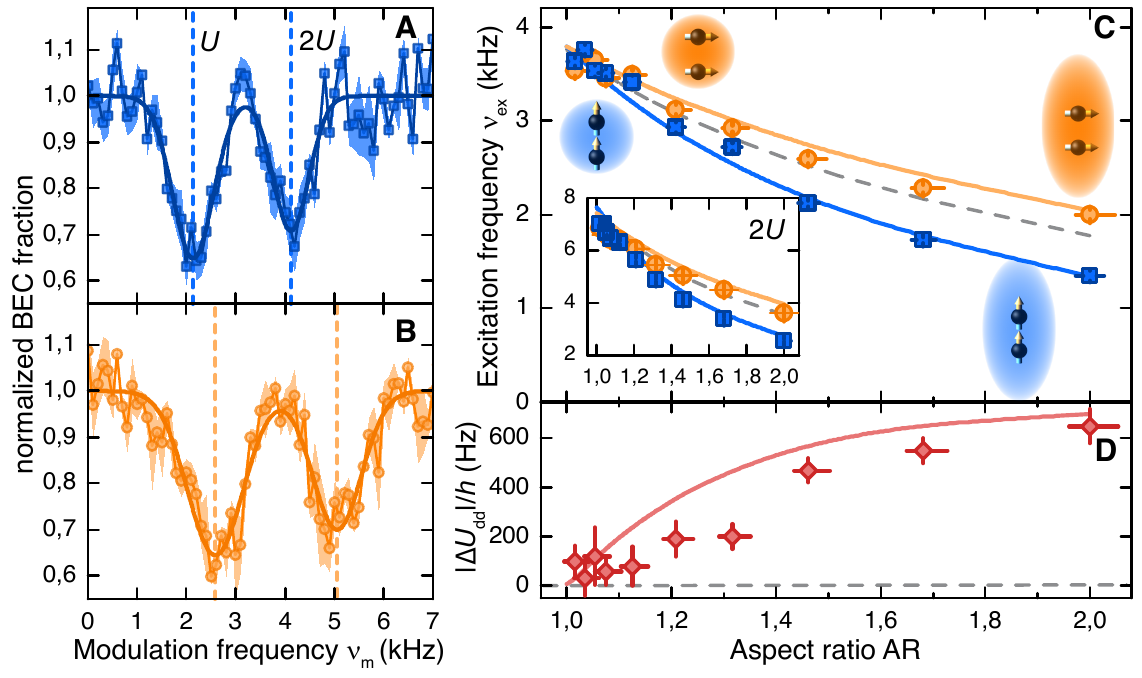}
	\caption{(color online) Measurement of the onsite interactions. (A and B) Excitation spectrum of the MI state for dipole orientations $\theta=\,0^\circ$ (A) and $\theta=\,90^\circ$ (B). The modulation spectroscopy is performed at $(s_{x},s_{y},s_{z})\,=\,(15,15,52.5)$, corresponding to ${\mathrm{AR}}\approx1.46$ and the remaining BEC fraction is measured after ramping down the lattice depths to zero. From a double Gaussian fit to the data (solid line) we extract the resonant excitation frequency $\nu_{\mathrm{ex}}$ for the $U$ and $2U$ feature. (C) $\nu_{\mathrm{ex}}$ for the loss feature at $U$ and $2U$ (inset) as a function of ${\rm AR}$ for $\theta\,=\,0^\circ$ (squares) and $\theta\,=\,90^\circ$ (circles). (D) Difference in the energy gap relative to the two dipole orientations, $|\Delta U_{\mathrm{dd}}|$, as a function of $\rm{AR}$. The error bars for all figures are the sum of the SEM and systematic errors (Supplementary Materials). The theoretical model (solid lines) reproduces very well the experimental data (C and D) and it also includes the effect of the NNI, which shifts the excitation frequency by up to $3\,\%$. For completeness, calculations accounting only for the isotropic (contact) interaction are shown (dashed lines).}
	\label{fig:fig_2}
\end{figure*}

We first investigate the impact of the DDI on the onsite interaction ($U_{\mathrm{dd}}$) by performing spectroscopic measurements. We prepare our system deep in the MI phase and probe the energy gap in the excitation spectrum for different dipole orientations. This energy gap, associated to particle-hole excitations, is $U$ for atoms in singly- or doubly-occupied Mott shells and $2U$ at the border between the two shells (Supplementary Materials)\,\cite{Jaksch1998cba}. We excite the MI by applying a sinusoidal modulation of frequency $\nu_{\mathrm{ex}}$ on the amplitude of the $x$-lattice beam \cite{Stoeferle2004,Kollath2006sou,Clark2006sot}. When $h\nu_{\mathrm{ex}}$ matches $U$ or $2U$, we observe a resonant depletion of the condensate. We perform the measurement for $\theta\,=\,0^\circ$ and $\theta\,=\,90^\circ$ (Fig.\,2A and B) and observe that the resonance positions clearly depend on the dipole orientation, consistent with our expectation.

To further explore this effect, we repeat the measurement for different values of the $\mathrm{AR}$ (Fig.\,2C). For the spherical case ($\mathrm{AR}=1$), we observe that the excitation gap looses its angle dependence showing that $U_{\mathrm{dd}}$ averages to zero\,\cite{Wall2013}. As the spatial distribution is deformed towards larger $\mathrm{AR}$, we find a clear deviation from the purely contact-interaction case (dashed lines), with a smaller energy gap for dipoles at $\theta\,=\,0^\circ$, and a larger one for dipoles at $\theta\,=\,90^\circ$.
Our measurement shows that $U_{\rm dd}$ plays a fundamental role in the stability of the MI phase: it can either protect the MI phase for the dominantly repulsive DDI ($\theta\,=\,90^\circ$) or make it more susceptible to excitations for the dominantly attractive case ($\theta\,=\,0^\circ$).
The energy difference between the two dipole configurations $|\Delta U_{\mathrm{dd}}|$ is shown in Fig.\,2D. The observed angle dependence is well described within our eBH model (Fig.\,2, C and D, solid lines), with $a_{\rm s}$ the only fit parameter. The derived value for $a_{\mathrm{s}}\,=\,137(1)\,\rm a_0$, with $\rm a_0$ being the Bohr radius, is consistent with previous measurements based on thermalization experiments\,\cite{Aikawa2012bec}.

\begin{figure*}[t]
	\includegraphics[width=1.5\columnwidth] {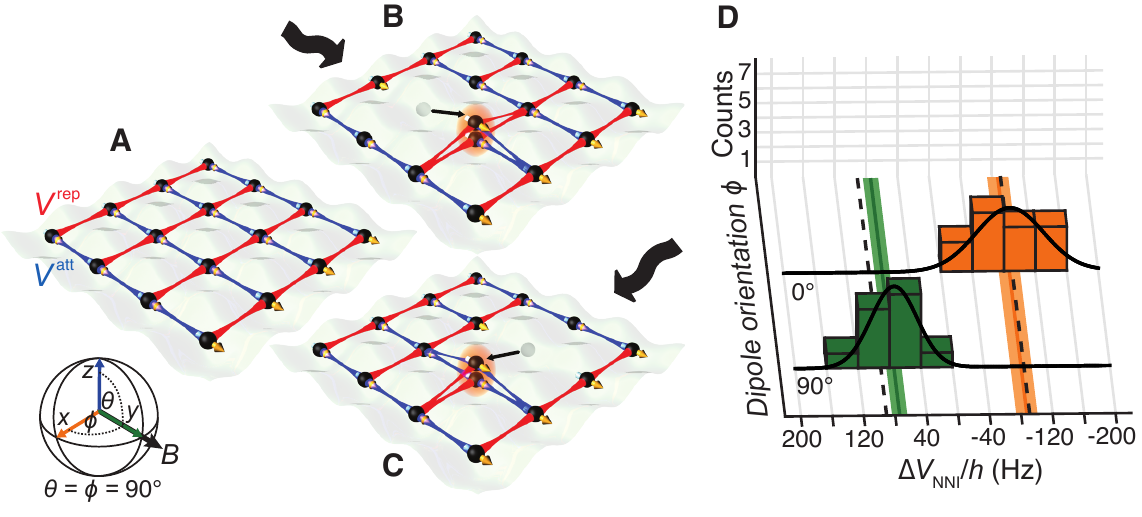}
	\caption{(color online) Nearest neighbor interactions. (A) Initial system in the MI regime with dipole orientation $\theta\,=\,90^\circ$, $\phi\,=\,90^\circ$. Driving excitations along $y$ (B) or $x$ (C) leads to two different particle-hole energy gaps $U-V^{\rm att}$ and $U-V^{\rm rep}$. Here, $V^{\rm att}=V_{ij}$ with in-plane head-to-tail dipole orientations and $V^{\rm rep}=V_{ij}$ for the in-plane side-by-side orientation. The difference between the two resonant energies $\Delta V_{\rm NNI}$, here equals to $-V^{\rm att} + V^{\rm rep}$, reveals the NNI. (D) Histogram of $\Delta V_{\rm NNI}$ for two dipole orientations $\phi\,=\,0^\circ$ and $\phi\,=\,90^\circ$. The solid lines in front of the histograms are the normal distributions of the corresponding data. On the bottom plane the two dashed lines show the theoretical expectation values, while the solid lines show the corresponding measured value with the shaded areas indicating the SEM.}
	\label{fig:fig_3}
\end{figure*}

In principle, the energy gap in the MI phase also depends on the NNI between atoms occupying adjacent lattice sites. However, in the above described measurement its influence is veiled by the much larger $U_{\rm dd}$. To demonstrate the presence of the NNI, we design a dedicated measurement scheme, which allows to isolate it from the other terms of the eBH. The measurement is based on modulation spectroscopy in the 2D short-spacing lattice plane ($xy$-plane), where the NNI is stronger (Fig.\,3A). The key idea is the following: For a system with only onsite interactions the energy gap associated with the particle-hole excitation does not depend on the direction of excitation, i.e. on the direction of the modulated beam. In contrast, a system including anisotropic NNI will exhibit a modification of the energy gap according to the excitation direction as the energy gap equals $U-V_{ij}$ for excitations along the bond direction $ij$. Hence the difference between the two resonance frequencies measured by modulating $s_x$ and $s_y$, denoted as $\Delta V_{\rm NNI}/h$, directly reveals the existence of the NNI as the onsite contribution cancels. Our scheme is illustrated in Fig.\,3, A to C, for the case $\theta=90^\circ,\phi=90^\circ$. Here, one bond of attractive (repulsive) NNI with energy $V^{\rm att}$ ($V^{\rm rep}$) gets destroyed during the excitation along (perpendicular to) the dipole orientation such that $\Delta V_{\rm NNI}= -V^{\rm att} + V^{\rm rep}$.
Our measurement for two dipole orientations in the plane with $\phi\,=\,90^\circ$ ($\,0^\circ$) give $\Delta V_{\rm NNI}/h\,=\,+74(10)\,$Hz ($\,-87(14)\,$Hz).  Remarkably, $|\Delta V_{\rm NNI}|$ is similar for both values of $\phi$ as expected from the symmetry between these two configurations
and is close to the theoretical expectation $h\times 91\,$Hz, as shown in Fig.\,3D. 
This set of measurement provides the first observation of the NNI predicted by the eBH model.

\begin{figure*}[t]
	\includegraphics[width=1.5\columnwidth] {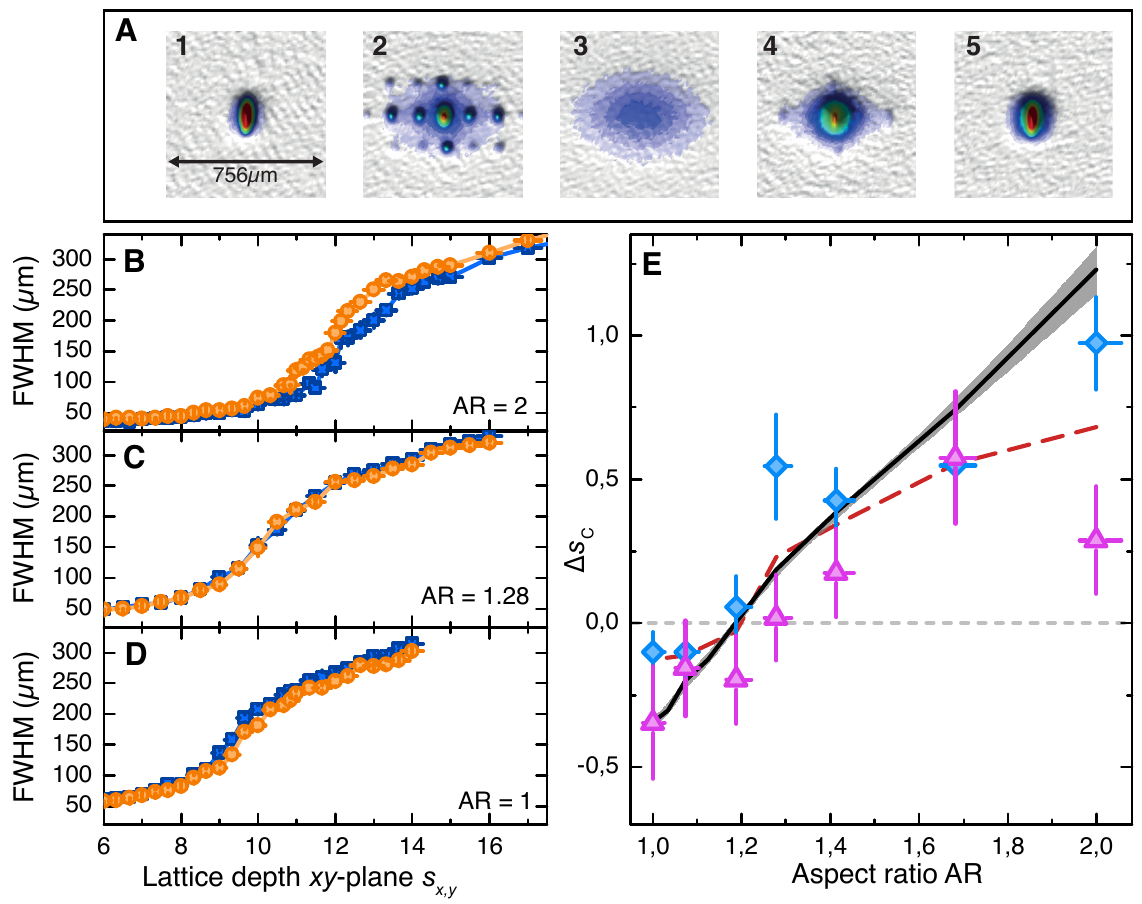}
	\caption{(color online) Superfluid-to-Mott-insulator transition. (A) Time-off-flight absorption images of the atomic cloud taken $27\,\rm{ms}$ after a sudden release from the 3D lattice with $s_{x,y,z}\,=\,s$ during ramp up $s=0$, $s=10$, $s=22$ (A, 1 to 3) and during ramp down $s=4$, $s=40$ (A, 4 to 5). (B to D) The width of the central interference peak is plotted as a function of the lattice depth in the $xy$-plane for $\textrm{AR}\,=\,2, 1.28, 1$, with dipoles oriented along $\theta\,=\,0^\circ$ (squares) and $\theta\,=\,90^\circ$ (circles). We extract the phase transition point $s_{\textrm{c}}$ for each orientation and AR via a double-line fit and the visibility (Supplementary Materials). (E) shows $\Delta s_{\rm{c}} = s_{\textrm{c}}(0^\circ)-s_{\rm{c}}(90^\circ)$ as a function of the $\rm{AR}$. The data from the visibility (diamonds) and the double-line fit (triangles) shows similar results. The dashed red line is the weighted mean of the two methods. The solid black line is the theoretical calculation from a mean field approximation (Supplementary Materials). The dotted line shows the expectation without any DDI.}
	\label{fig:fig_4}
\end{figure*}

Finally, we study the effect of the anisotropic DDI on the many-body phase transition from a SF to a MI phase (Fig.\,4A). This phase transition is a result of the competition between interactions and tunneling \cite{greiner2002quantum}, therefore also revealing the influence of the DAT \cite{note6}. We probe the atomic interference patterns of the expanding cloud after a sudden release from the 3D lattice for different dipole orientations ($\theta\,=\,0^\circ$ and $90^\circ$). For increasing lattice depths the system enters the MI phase, as shown by the disappearance of the interference pattern and the increase of the full width at half maximum (FWHM) of the central interference peak (Fig. 4, B to D). We clearly observe a shift of the phase transition as a function of the lattice depth depending on $\theta$ and $\mathrm{AR}$. For large $\rm ARs$ (Fig. 4B), the MI phase is favored for $\theta=90^\circ$ with respect to $\theta\,=\,0^\circ$ as expected from the angle dependence of $U_{\rm dd}$. For the spherical case ($\mathrm{AR}\,=\,1$) (Fig.\,4D), contrary to the naive expectation, we also observe the shift of the phase transition, which is now inverted, with the MI phase favored for $\theta\,=\,0^\circ$, as compared to the case of large $\mathrm{AR}$. Instead, we find that the shift of the phase transition vanishes at $\mathrm{AR}\,\approx\,1.2$ (Fig.\,4C).
This behavior is a direct consequence of the action of the anisotropic DAT term $\Delta J_{ij,{\rm dd}}$ in the eBH. For a more quantitative analysis, we systematically study the phase transition as a function of $\mathrm{AR}$ for $\theta\,=\,0^\circ$ and $\theta\,=\,90^\circ$. We extract the critical value of the lattice depth, $s_c(\theta)$, which defines the onset of the SF-to-MI phase transition (Supplementary Materials). As shown in Fig.\,4E, the difference $\Delta s_{\rm{c}} = s_{\rm{c}}(0^\circ)-s_{\rm{c}}(90^\circ)$ decreases when lowering the $\mathrm{AR}$, crosses zero at $\mathrm{AR}\,\approx\,1.2$, and eventually become negative at even smaller $\mathrm{AR}$. This behavior is very well reproduced by our calculations using the full eBH model within a mean-field approximation (Supplementary Materials), highlighting the importance of DAT in the many-body system dynamics.

Quantum degenerate gases of magnetic Lanthanide atoms in optical lattices offer a new avenue to access the physics of strongly correlated systems for both bosonic and fermionic Hubbard dynamics in the presence of dipolar interactions, while building on the well-developed toolbox to prepare ultracold dense samples, and manipulate and measure these atomic gases. We have realized the extended Bose-Hubbard Hamiltonian with anisotropic onsite and offsite interactions, which reveal themselves in the excitation spectrum and in the many-body dynamics of the system. Our results show how to control the Hamiltonian terms with the dipole orientation and accomplish the long-awaited observation of NNI in Hubbard dynamics. An outstanding challenge for future experiments is the preparation of exotic quantum phases for bosons and fermions: an example is provided by stripe phases due to NNI, which become accessible with present interaction strengths at temperatures in the few nK regime (Supplementary Materials). Dipolar interactions can be increased by working with Feshbach molecules of magnetic Lanthanides, essentially doubling the magnetic dipole moment \cite{frisch2015ultracold}. These opportunities offered by Lanthanides to access the multitude of many-body phases predicted for dipolar quantum matter are complemented by the remarkable experimental developments with heteronuclear molecules and Rydberg atoms \cite{Baranov2012}.

\begin{acknowledgments}
We thank F. Meinert and H.-C.-N\"agerl for fruitful discussions. The Innsbruck experimental group is supported by the Austrian Ministry of Science and Research (BMWF) and the Austrian Science Fund (FWF) through a START grant under project Y479-N20 and by the European Research Council (ERC) under project 259435. K.A. has been supported within the Lise-Meitner program of the FWF. The Innsbruck theory group is supported by the SFB FoQuS, by the ERC Synergy Grant UQUAM, and by the EU FET Proactive Initiative SIQS. 
\end{acknowledgments}

\appendix

\section*{Supplementary Materials}

\subsection*{BEC production}
We create a BEC of about $1.5\times 10^5$ $^{168}$Er atoms by means of evaporative cooling in a crossed optical dipole trap (ODT)\,\cite{Aikawa2012bec}. The cloud has typically a BEC fraction above $80\,\%$, which is extracted by a two-dimensional bimodal fit to an absorption image of the atomic cloud after a time-of-flight (TOF) of $27\,\rm ms$\,\cite{Aikawa2012bec}. The cloud temperature is estimated to be about $70\,\rm nK$. The ODT is operated at $1064\,\rm nm$ and is created by two beams, one propagating horizontally and one vertically. The beams cross at their respective focal points. The elliptic horizontal beam has a vertical (horizontal) waist of about $18\,\mathrm{\mu m}$ ($117\,\mathrm{\mu m}$) and the elliptic vertical beam has a waist of about $55\,\mathrm{\mu m}$ ($110\,\mathrm{\mu m}$) along (perpendicular to) the axis of the horizontal beam. The measured trap frequencies are $(\omega_x, \omega_y, \omega_z)\,=\,2\pi\times(29.0(6), 22.2(4), 165.2(5))\,\rm Hz$. We observe a lifetime of the trapped cloud of about $10\,\rm s$. 

The atomic cloud is spin-polarized in the lowest Zeeman sublevel $(J\,=\,6,m_J\,=\,-6)$, where $J$ denotes the total angular momentum quantum number and $m_J$ is its projection along the quantization axis. The spin polarization already occurs in the magneto-optical trap\,\cite{Frisch2012nlm} and is maintained in the ODT by applying a bias magnetic field with a fixed value of $0.40(1)\,\rm G$. As discussed below, the magnitude of this field is kept constant for all the experiments, whereas its orientation is varied to set the desired dipole orientation.

\subsection*{3D lattice setup}
We describe the 3D lattice setup in the coordinate system given by the two horizontal lattice beams denoting the $x$ and $y$-axis and the direction of gravity giving the $z$-axis (inset, Fig. 1A).
The horizontal lattice beams are created by two retroreflected beams with a waist of about $160\,\rm \mu m$ and a wavelength $\lambda_{x}\,=\,\lambda_{y}\,=\,532\,$nm. The vertical lattice beam has a waist of about $300\,\rm \mu m$ and a wavelength $\lambda_{z}\,=\,1064\,$nm. The resulting 3D optical lattice is given by $V(x,y,z)=V_{x}\cos^{2}(k_{x}x)+V_{y}\cos ^{2}(k_{y}y)+V_{z}\cos ^{2}(k_{z}z)$, where $V_{i}$ is the lattice depth in the $i$-direction and $k_{i}\,=\,2\pi/\lambda_{i}$ the corresponding lattice wavevector with $i\,=\,(x,y,z)$. Because of the different wavelengths, the atoms experience different recoil energies $\mathrm{E}_{\mathrm{R},i}$ in the $xy$-plane with respect to the vertical direction. The recoil energies given by $\mathrm{E}_{\mathrm{R},i}\,=\,h^2/(2m\lambda_i^2)$ are $\mathrm{E}_{\mathrm{R},x}\,=\,\mathrm{E}_{\mathrm{R},y}\,=\,h\times4.2\,$kHz and $\mathrm{E}_{\mathrm{R},z}\,=\,h\times1.05\,$kHz. Here, $h$ is the Planck constant and $m$ the mass of the Er atom. For convenience, we give the lattice depth in units of the corresponding recoil energy $s_i\,=\,V_i/\mathrm{E}_{\mathrm{R},i}$. The maximum lattice depth we can achieve is $(s_x,s_y,s_z)\,=\,(30,30,220)$. Because of the Gaussian profile of the lattice beams the atoms experience an additional harmonic confinement. At a typical 3D lattice depth of $(s_x,s_y,s_z)\,=\,(20,20,20)$ we measure $(\omega_x, \omega_y, \omega_z)\,=\,2\pi\times(34(1), 31(1), 43(1))\,\rm Hz$.

We note that the vertical lattice beam is tilted from the vertical axis by $\theta\,=\,10(2)^\circ$ and has an azimuthal angle of $\phi\,=\,5(5)^\circ$. This has two consequences: (a)  The lattice spacings $d_x$ and $d_z$ are modified to $d_x\,=\,270(2)\,$nm and $d_z\,=\,540(4)\, \rm nm$ with respect to the $\lambda/2$ case and (b) the tilt of the wavefront of the vertical lattice beam gives rise to an additional potential difference between neighboring lattice sites along $x$ of $200(40)\,\rm Hz$ due to gravity. While (b) only leads to a broadening of the excitation resonances in the modulation spectroscopy measurements, (a) could in principle change the values of the eBH terms. Therefore, we recalculate them considering our effective lattice spacings for a typical experimental condition of $(s_x,s_y,s_z)\,=\,(15,15,15)$. We find that the isotropic terms are reduced by $3\,\%$ while the anisotropic terms can differ between $2\textrm{-}6\,\%$, depending on the dipole orientation and the direction of the observed process (see Table 1). This gives rise to a downshift of the phase transition point $s_{\textrm{c}}$ of about $1\,\%$ for both $\theta\,=\,0^\circ$ and $\theta\,=\,90^\circ$. However, all these shifts are not resolvable within our statistical errors and can therefore safely be neglected.

\begin{table}[htb]
	\caption{Difference of the eBH terms between the $\lambda/2$-spacing and the actual spacing given in percentage of the $\lambda/2$-case for three dipole orientations ($\theta\,=\,90^\circ$, $\phi\,=\,0^\circ$), ($\theta\,=\,90^\circ$, $\phi\,=\,0^\circ$), and $\theta\,=\,0^\circ$.}
	\centering
	\begin{tabular}{|c||c|c|c|}
		\hline  & $\theta\,=\,90^\circ$, $\phi\,=\,0^\circ$ & $\theta\,=\,90^\circ$, $\phi\,=\,0^\circ$ & $\theta\,=\,0^\circ$ \\ 
		\hline\hline $U_{\rm s}$ &  \multicolumn{3}{c|}{$3\,\%$} \\ 
		\hline $J_{ij=x},J_{ij=z}$ & \multicolumn{3}{c|}{$3\,\%$}  \\ 
		\hline $J_{ij=y}$ & \multicolumn{3}{c|}{$0\,\%$}  \\
		\hline\hline $U_{\rm dd}$ & $6\,\%$ & $-2\,\%$ & $2\,\%$ \\ 
		\hline $U$ & $3\,\%$ & $-2\,\%$ & $3\,\%$ \\ 
		\hline $\Delta V_{\rm NNI}$ & $3\,\%$ & $2\,\%$ & - \\ 
		\hline $\Delta J_{ij=x}$ & $3\,\%$ & $2\,\%$ & $3\,\%$ \\ 
		\hline $\Delta J_{ij=y}$ & $4\,\%$ & $2\,\%$ & $3\,\%$ \\ 
		\hline $\Delta J_{ij=z}$ & $3\,\%$ & $2\,\%$ & $3\,\%$ \\ 
		\hline 
	\end{tabular} 
\end{table}

\subsection*{Lattice depth calibration and onsite aspect ratio}
To calibrate the depths of the horizontal lattice beams we use the standard Kapitza-Dirac diffraction method \cite{Gould1986doa}. For the vertical lattice we use the technique of parametric heating, in which the atoms are excited from the first to the third lattice band \cite{Denschlag2002abe,Morsch2006dob}.
With these methods, we extract the lattice depths with an uncertainty of up to $4\,\%$.

The onsite $\rm AR$ is defined in terms of a Gaussian approximation to the corresponding Wannier function: $\mathrm{AR}=l_z/l_{x,y}$, where $l_{x,y}=d_{x,y}/(\pi s_{x,y}^{1/4})$ and $l_z=d_z/(\pi s_z^{1/4})$ are the harmonic oscillator lengths associated to the lattice beams along $x,y$ and $z$ respectively (Note that we use $s_x=s_y$ in our measurements). The uncertainty of the $\rm AR$ results from the uncertainty of the lattice depths and is about $1\,\%$.

Because of the non $S$-state character of Er atoms in their electronic ground state, the atomic polarizability of Er has a tensorial contribution, which is about $3\,\%$ of the scalar one for an off-resonant trapping light \cite{Lepers2014aot}. In our system this effect gives rise to a different lattice depth depending on the dipole orientation. We carefully studied this effect by calibrating each lattice beam for both orientations, dipoles aligned parallel or orthogonal to the lattice beam. Our measurements reveal that a parallel orientation gives an up to $4\,\%$ deeper confinement compared to the orthogonal orientation. 
For simplicity, we account for this effect only by a systematic error in the $\mathrm{AR}$, leading for instance to the asymmetric error bars in Fig.\,2(C and D), and Fig.\,4E.

\subsection*{Loading of the 3D lattice}
For our experiments the atoms are adiabatically loaded to the 3D lattice by an exponential ramp to the final value within $150\,\rm ms$, during which the vertical ODT is linearly lowered to zero. 
To perform modulation spectroscopy, the horizontal ODT is switched off within $1\,\rm ms$ after the loading.
For the measurement of the BEC depletion, we exactly reverse the described process.
In the MI phase we estimate a central density of two atoms per lattice site. The external harmonic confinement leads to a density distribution with a central doubly occupied Mott shell, consisting of up to 40\,\% of the total atoms, surrounded by a singly occupied shell. The external harmonic confinement is given by the sum of the ODT potentials and the Gaussian profiles of the lattice beams during the lattice loading.
For our typical lattice depth condition $(s_x,s_y,s_z)\,=\,(20,20,20)$, the lifetime of the atomic sample in the lattice is $5(1)\,\rm s$. In addition, we observe a heating, which leads to a full depletion of the recovered BEC for a holding time in the lattice of about $1\,\rm s$. The origin of this heating is not fully understood and might be due to frequency fluctuations of the $532\,\rm nm$ laser source.

\subsection*{Control of the dipole orientation}
The dipole orientation follows the direction of the magnetic field, which we control using three pairs of independent coils oriented perpendicular to each other. Each pair of coils is independently calibrated by performing radio-frequency spectroscopy, where resonant excitations to higher Zeeman sublevels can be used as a measure of the actual magnetic field at the position of the atoms. The dipole orientation can be changed from $\theta\,=\,0^\circ$ to $\theta\,=\,90^\circ$ and for any value of $\phi$. Noise of the ambient magnetic field leads to fluctuations of the absolute angles $\theta$ and $\phi$ by about $1^\circ$ around their set values. 
During the evaporative cooling sequence the dipoles are aligned at $\theta\,=\,0^\circ$. Before loading the atoms into the 3D lattice, the dipole orientation is changed to the desired value in $38\,\rm ms$, while the magnetic-field magnitude is kept constant. After the release of the atoms from the trap, the magnetic field is rotated towards the imaging direction ($\theta\,=\,90^\circ$ and $\phi\,=\,160^\circ$) to perform standard absorption imaging\,\cite{Aikawa2012bec}.

\subsection*{Modulation spectroscopy in the MI}
To probe the excitation gap in the MI we use a modulation spectroscopy technique\,\cite{Kollath2006sou,Stoeferle2004}. We sinusoidally modulate the power of one horizontal lattice beam with a typical total amplitude between $30\,\%$ and $40\,\%$, and a modulation time between $50\,\rm ms$ and $100\,\rm ms$. With this method, we resonantly create particle-hole excitations in the system\,\cite{Clark2006sot}. These excitations manifest themselves as a resonant depletion of the recovered BEC because of the extra energy stored in the system. We record the remaining BEC fraction after ramping down the lattice  as a function of the modulation frequency. The resulting loss spectrum is then fitted with a double-Gaussian function, whose centers give the excitation frequencies. The typical FWHM of the resonant loss features is $1\,\rm kHz$ for excitations using the $x$-lattice beam and $0.8\,\rm kHz$ for the $y$-lattice beam. The width is mainly determined by the external harmonic confinement. We note that the difference in width between the two excitation directions is due to the tilt of the vertical lattice beam as discussed above.

\begin{figure}[t]
	\includegraphics[width=0.8\columnwidth] {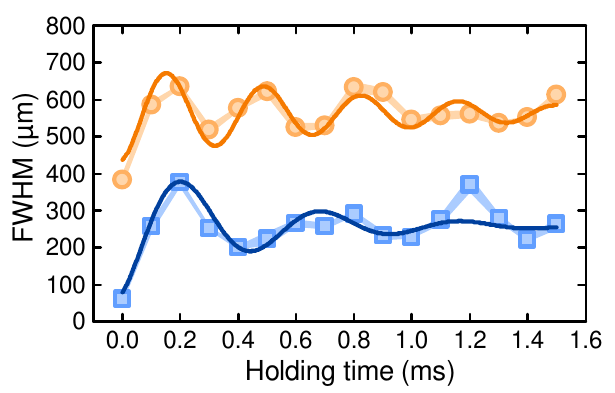}
	\caption{(color online) Measurement of $U$ by the collapse-and-revival technique. The FWHM of the central peak of the interference pattern is monitored as a function of the holding time after a sudden quench from the SF to the MI phase for an initial dipole orientation of $\theta\,=\,0^\circ$ (squares) and $\theta\,=\,90^\circ$ (circles). The latter measurement is vertically offset by $300\,\rm \mu m$ for a better visualization of the two data sets. Each point is obtained by two to four independent measurements and the shaded region indicates the SEM. The solid lines are the fits of a damped sine to the data, used to extract the oscillation frequency and hence $U$.}
	\label{fig:fig_S1}
\end{figure}

We also measure the onsite interaction by using an alternative method, known as the collapse-and-revival technique\,\cite{Greiner2002car}. Here, we first prepare the system at the onset of the SF-to-MI transition with a lattice depth of $(s_x,s_y,s_z)\,=\,(10,10,10)$ and we then suddenly quench the system to $(s_x,s_y,s_z)\,=\,(20,20,40)$ within $5\,\rm \mu s$. As a result of the quench the system oscillates between the MI and the SF phase. Figure S1 shows the evolution of the FWHM of the central interference peak as a function of the holding time after the quench for two different dipole orientations $\theta\,=\,0^\circ$ and $\theta\,=\,90^\circ$. We observe up to four collapses and revivals and extract the onsite interaction from the oscillation frequency. For $\theta\,=\,0^\circ$ ($\theta\,=\,90^\circ$) we measure a frequency of $2.07(16)\,\rm kHz$ ($2.98(5)\,\rm kHz$), which are consistent with the value of $2.15(3)\,\rm kHz$ ($2.77(3)\,\rm kHz$) obtained with the modulation spectroscopy technique.

\subsection*{Analysis of the NNI}

To derive the NNI we perform a differential measurement based on modulation spectroscopy, in which the orientation of dipoles is fixed but the direction of excitation is changed between the horizontal lattice axes $x$ and $y$. To explain the amount of energy needed to drive a particle-hole excitation we consider the situation where the dipoles are aligned with angles $\theta\,=\,90^\circ$ and $\phi=90^\circ$, as also illustrated in Fig.\,3 (see main manuscript). Here we denote $V^{\rm att}$ ($V^{\rm rep}$) the attractive (repulsive) value of $V_{ij}$ for the bond direction $y$ ($x$). At the starting configuration (Fig.\,3A) the total energy is $E_{\rm A}\,=\,12 V^{\rm att}+12 V^{\rm rep}$. For an excitation along the $y$-axis the final energy of this configuration reads as $E_{\rm B}\,=\,U+11 V^{\rm att}+12 V^{\rm rep}$, while for an $x$-excitation it is $E_{\rm C}\,=\,U+12 V^{\rm att}+11 V^{\rm rep}$. From this consideration it becomes clear that the difference in energy $E_{\rm B}-E_{\rm C}\,=\,-V^{\rm att}+V^{\rm rep}\,=\,\Delta V_{\rm NNI}$ purely reveals the NNI. 

Analogously the same consideration can be applied for an initial dipole orientation of $\theta\,=\,90^\circ$ and $\phi=0^\circ$ leading to $\Delta V_{\rm NNI}=-V^{\rm rep}+V^{\rm att}$.
From the theory we expect $V^{\rm rep}/h=31.5\,\rm Hz$, $V^{\rm att}/h=-59.5\,\rm Hz$ and thus $|\Delta V_{\rm NNI}|/h=91\,\rm Hz$. Including the corrections arising from the modification of the lattice spacings due to the tilt of the vertical lattice beam (see above) 
$|\Delta V_{\rm NNI}|/h$ changes to $89\,\rm Hz$, even closer to our measured values.

\begin{figure*}[t]
	\includegraphics[width=1.5\columnwidth] {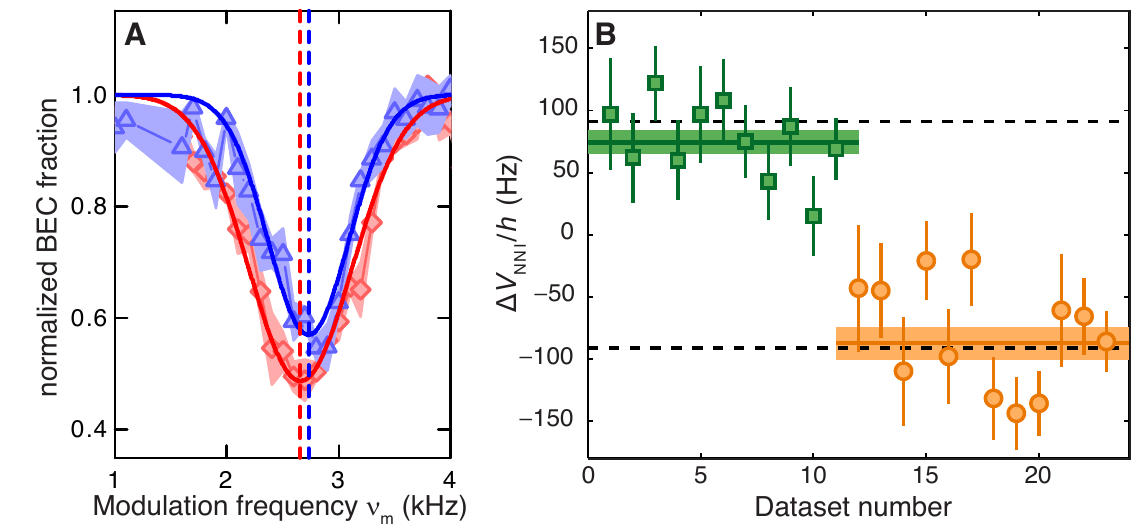}
	\caption{(color online) Measurements for the NNI. (A) Excitation spectrum with modulation along $x$ (diamonds) and $y$ (triangles) with $\theta\,=\,90^\circ$ and $\phi\,=\,90^\circ$, forming one differential measurement. Each point is the average of about $5$ independent measurements and the shaded region indicates the SEM. The solid lines are weighted Gaussian fits to the data. The dashed lines indicate the obtained resonance frequencies. (B) Set of differential measurements as presented in (A) for dipoles aligned with  $\theta\,=\,90^\circ$ and $\phi\,=\,90^\circ$ (squares),  $\theta\,=\,90^\circ$ and $\phi\,=\,0^\circ$ (circles). 
		The solid lines are weighted fits with the shaded region being the SEM. The dashed lines are the theoretical expectations.}
	\label{fig:fig_S2}
\end{figure*}

In Fig.\,S2A we show two excitation spectra obtained using the method described above described. The difference between the centers of the Gaussian fits to the data is found to be $72\rm(30)\,Hz$ and corresponds to one data point of Fig.\,S2B, where all taken measurements are summarized. We believe that the fluctuation of $\Delta V_{\rm NNI}$ along the data sets is mainly caused by relative drifts of the lattice depths during a differential measurement. We carefully check for systematic errors on $\Delta V_{\rm NNI}$ using different initial lattice depths or atom numbers, but do not find an effect within our measurement resolution. The used lattice depths are $(s_x,s_y,s_z)\,=\,(15, 15, 30),(14,18,30),$ and $(20,20,40)$. 
The different depths can slightly modify $\Delta V_{\rm NNI}$ by maximum $2\,\%$ which is not resolvable within our error bar.

\subsection*{Analysis of the SF-to-MI transition point}

The critical value of the SF-to-MI transition, $s_{\textrm{c}}$, depends on the ratio of the total onsite interaction to the total tunneling rate. In presence of dipole-dipole interaction (DDI), both terms depend on the dipole orientations, imprinting an angle dependence on the phase transition point $s_{\textrm{c}}(\theta)$, defined by the value of the horizontal lattice depth $s_{x,y}\,=\,s_x\,=\,s_y$. We study the phase transition for  $\theta\,=\,0^\circ$ and $\theta\,=\,90^\circ$ and extract the difference in the critical point  $\Delta s_{\textrm{c}}\,=\,s_{\textrm{c}}(0^\circ)-s_{\textrm{c}}(90^\circ)$. In particular, we ramp up simultaneously the three lattice beams in $150\,\rm ms$ while lowering the vertical ODT to zero. We then suddenly switch off all beams, let the cloud expand for $27\,\rm ms$ and perform standard absorption imaging. We repeat this cycle for various final lattice depths $s_{x,y}$ and various values of the $\rm{AR}$.

\begin{figure*}[t]
	\includegraphics[width=1.5\columnwidth] {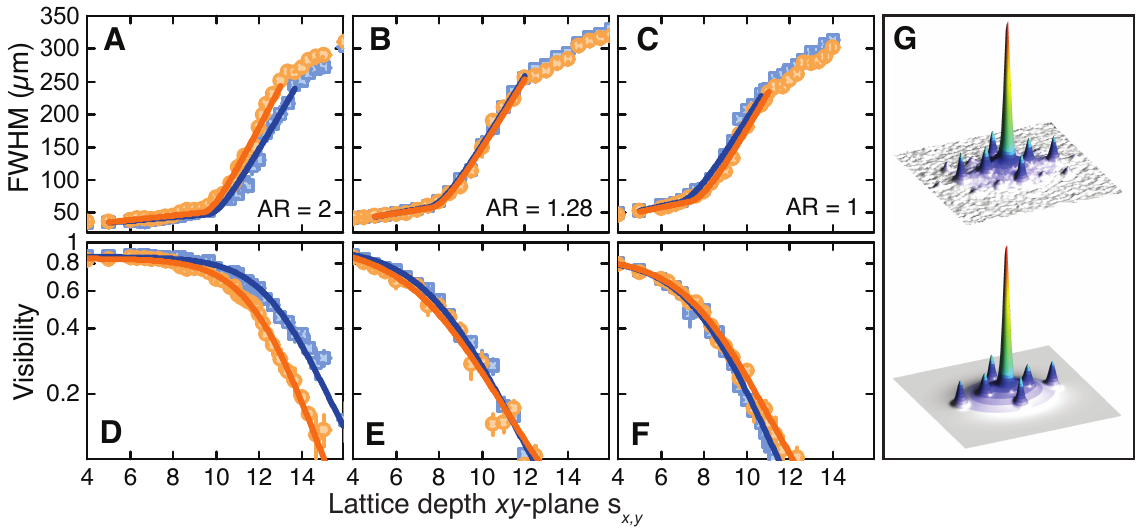}
	\caption{(color online) Derivation of the phase transition point $s_{\textrm{c}}$. (A to F) The FWHM of the central interference peak and the visibility are displayed as a function of the final $s_{x,y}$ lattice depths for $\textrm{AR}\,=\,2$  (A and D), $1.28$ (B and E), and $1$ (C and F), for dipole orientations $\theta\,=\,0^\circ$ (squares) and $\theta\,=\,90^\circ$ (circles). The solid lines show the corresponding fitting functions (see text). Each point is obtained from about $10$ independent measurements. (G) Example of density distribution after TOF (top) and the corresponding 2D fit (bottom) used for the derivation of the visibility.}
	\label{fig:fig_S3}
\end{figure*}

To extract $s_{\textrm{c}}$ we use two methods. The first method (a) analyses the increase of the FWHM of the central interference peak as a function of $s_{x,y}$ (Figure S3, A to C). In general we observe a smooth transition from the SF phase (small FWHM) to the MI phase (large FWHM) as expected for a trapped system with a spatially depending density. A weighted fitting function consisting of two smoothly connected lines is used to extract the critical depth. The region for fitting starts at $s_{x,y}\,=\,5$ and goes up to a maximum FWHM of $250\,\rm \mu m$ for all data. The point where the two lines are crossing is interpreted as the phase transition point $s_{\textrm{c}}$. We carefully check the dependence of the chosen boundaries of the fit region and do not find a significant influence on the qualitative behavior of $\Delta s_{\textrm{c}}$ . However, quantitatively $\Delta s_{\textrm{c}}$ can vary by up to $\pm 0.2$.

The second method (b) analyses the visibility. Figure S3, D to E, shows the extracted visibility data for the same data as in (a). The visibility is calculated from a two-dimensional fit consisting of seven gaussians for the central and first-order interference peaks and one broad gaussian for the incoherent background to the obtained interference pattern (see Fig.\,S2G). The visibility is defined as $\mathcal{V}=A/(A+B)$, where  $A$ stands for the mean amplitude of the first-order interference peaks and $B$ for the mean value of the incoherent background computed at the positions of the interference peaks. 
We extract $\mathcal{V}$ as a function of $s_{x,y}$ and fit the whole dataset by the phenomenological function $\mathcal{V}(s_{x,y})\,=\,C/(1+\mathrm{exp}(\alpha(s_{x,y}-s_{\textrm{c}})))-\mathcal{V}_0$ (adapted from\,\cite{Ospelkaus2006lob}). Here $C$, $\alpha$, $s_{\textrm{c}}$, and $\mathcal{V}_0$ are fitting parameters, where $s_{\textrm{c}}$ corresponds to the phase transition point.

For both methods, at large $\rm ARs$, we observe a clear shift of $s_{\textrm{c}}$ toward higher values for $\theta\,=\,0^\circ$ compared to $\theta\,=\,90^\circ$ (Fig.\,S3, A and D).  
This shift gets reduced when lowering the $\rm AR$, vanishes around $\mathrm{AR} \approx 1.2$ (Fig.\,S3, B and E), and changes sign for $\mathrm{AR} = 1$ (Fig.\,S3, C and F).

\subsection*{Extended Bose-Hubbard model from microscopic Hamiltonian}

Here we present the details of derivation of the eBH model Eq.\,1 together
with the expressions for all its coefficients in terms of microscopic
parameters of the system (see, e.\,g.\,\cite{Mazzarella2006ebh}).

The microscopic Hamiltonian of the consider system of polarized (magnetic)
dipolar atoms has the form: 
\begin{equation}
\hat{H}_{\rm tot}=\hat{H}_{0}+\hat{H}_{\rm int},  \label{eq:total}
\end{equation}%
where the first term 
\begin{equation}
\hat{H}_{0}=\int d\mathbf{r}\Psi ^{\dag }(\mathbf{r})[-\frac{\hbar
	^{2}\nabla ^{2}}{2m}+V(\mathbf{r})]\Psi (\mathbf{r})  \label{H0}
\end{equation}%
is the single-particle Hamiltonian with $\Psi (\mathbf{r})$ being a boson
field operator, which describes the motion of an atom with the mass $m$ in
the optical-lattice potential $V(\mathbf{r})=V_{x}\cos
^{2}(k_{x}x)+V_{y}\cos ^{2}(k_{y}y)+V_{z}\cos ^{2}(k_{z}z)$, and the second
term 
\begin{equation}
\hat{H}_{\rm int}=\frac{1}{2}\int d\mathbf{r}\int d\mathbf{r}^{\prime }\Psi
^{\dag }(\mathbf{r})\Psi ^{\dag }(\mathbf{r^{\prime}})U(\mathbf{r}-\mathbf{r}^{\prime
})\Psi (\mathbf{r}^{\prime })\Psi (\mathbf{r})  \label{Hint}
\end{equation}%
corresponds to the interatomic interaction. In the considered case, the
interaction contains a short-range part, which can be modeled by a contact
potential with the $s$-wave scattering length $a_{\rm s}$, and the DDI (see, e.g. \cite{Yi2000tac})%
\begin{equation*}
U(\mathbf{r}-\mathbf{r}^{\prime })=\frac{4\pi \hbar a_{\rm s}}{m}\delta (\mathbf{%
	\ r}-\mathbf{r}^{\prime })+\frac{\mu _{0}\mu ^{2}}{4\pi }\frac{1-3\cos
	^{2}\theta _{\mathbf{r}-\mathbf{r}^{\prime }}}{|\mathbf{r}-\mathbf{r}
	^{\prime }|^{3}}
\end{equation*}%
with $\theta _{\mathbf{r}-\mathbf{r}^{\prime }}$ being the angle between the
relative position of two dipoles $\mathbf{r}-\mathbf{r}^{\prime }$ and their
polarization.

The Hamiltonian of Eq.\,3 determines the single-particle band structure,%
\begin{equation*}
\lbrack -\frac{\hbar ^{2}\nabla ^{2}}{2m}+V(\mathbf{r})]u_{\alpha \mathbf{p}%
}(\mathbf{r})=\varepsilon _{\alpha }(\mathbf{p})u_{\alpha \mathbf{p}}(%
\mathbf{r}),
\end{equation*}%
where $u_{\alpha \mathbf{p}}(\mathbf{r})$ is the Bloch wavefunction
corresponding to the band $\alpha $ and quasimomentum $\mathbf{p}$ from the
Brillouin zone (BZ), defined by $-\pi /d_{i}<p_{i}/\hbar \leq \pi /d_{i}$ with $%
d_{i}=\pi /k_{i}$ being the lattice spacing along the $i$-direction, $p_i$ the corresponding component of $\bf p$, and $%
\varepsilon _{\alpha }(\mathbf{p})$ is the corresponding energy. For our
purposes it is more convenient to work with Wannier functions $\phi
_{i,\alpha }(\mathbf{r})=\sum_{\mathbf{p}\in \mathrm{BZ}}\exp [-i\mathbf{p(r}%
-\mathbf{R}_{i}\mathbf{)}]u_{\alpha \mathbf{p}}(\mathbf{r})$, which are
localized at different sites $\mathbf{R}_{i}$ of the lattice and orthogonal
to each other with respect to both the lattice position $i$ and the band
index $\alpha $, $\int d\mathbf{r}\phi _{i,\alpha }^{\ast }(\mathbf{r})\phi
_{j,\beta }(\mathbf{r})=\delta _{ij}\delta _{\alpha \beta }$. Using these
functions as a single-particle basis in the bosonic field operator, $\Psi (%
\mathbf{r})=\sum_{i,\alpha }b_{i,\alpha }\phi _{i,\alpha }(\mathbf{r})$,
where $b_{i,\alpha }$ are the bosonic annihilation operators for particles
on the site $i$ in the band $\alpha $, we can rewrite the initial
Hamiltonian (\ref{H0}) in terms of the operators $b_{i,\alpha }$ and $%
b_{i,\alpha }^{\dag }$. To obtain the eBH model, we keep terms with
operators for the lowest energy band only. Note that this approximation is
legitimate because the interatomic interaction in our case is order of magnitude
less than the band gap such that the admixture of the higher bands can be
neglected. From the remaining terms we then neglect those which contain
square and higher power of exponentially small spatial overlaps of the
Wannier functions from different sites (see, \thinspace \cite%
{Mazzarella2006ebh} for details). Denoting the operators and the Wannier
functions for the lowest band as $b_{i}$, $b_{i}^{\dag }$ and $\phi _{i}(%
\mathbf{r})\,$, respectively, we obtain 
\begin{equation}
\begin{split}
H=& -\sum_{\langle ij\rangle }J_{ij}(b_{i}^{\dag }b_{j}+h.c)+\frac{U}{2}%
\sum_{i}n_{i}(n_{i}-1)+\sum_{\langle ij\rangle
}V_{ij}n_{i}n_{j}\\
& -\sum_{\langle ij\rangle }\Delta J_{ij}[b_{i}^{\dag
}b_{j}(n_{i}+n_{j}-1)+h.c]  \label{HeHm}
\end{split}
\end{equation}%
where $\langle ij\rangle $ denotes a pair of nearest-neighboring sites. The
first two terms in this expression correspond to the standard Hubbard model
with the single-particle hopping amplitude 
\begin{equation*}
J_{ij}=\int d\mathbf{r}\phi _{i}^{\ast }(\mathbf{r})[-\frac{\hbar ^{2}\nabla
	^{2}}{2m}+V(r)]\phi _{j}(\mathbf{r})
\end{equation*}%
and the onsite interaction $U=U_{\rm s}+U_{\rm dd}$, where $U_{\rm s}$ comes from the
contact interaction, 
\begin{equation*}
U_{\rm s}=\frac{4\pi \hbar a_{s}}{m}\int d\mathbf{r}\left\vert \phi _{i}(\mathbf{%
	\ r})\right\vert ^{4},
\end{equation*}%
and $U_{\rm dd}$ from the dipole-dipole one, 
\begin{equation*}
U_{\rm dd}=\frac{\mu _{0}\mu ^{2}}{4\pi }\int d\mathbf{r}\int d\mathbf{r^{\prime
	}}\left\vert \phi _{i}(\mathbf{r})\right\vert ^{2}\frac{1-3\cos ^{2}\theta _{%
	\mathbf{r}-\mathbf{r}^{\prime }}}{|\mathbf{r}-\mathbf{r}^{\prime }|^{3}}%
\left\vert \phi _{i}(\mathbf{r}^{\prime })\right\vert ^{2}.
\end{equation*}%
The third term in Eq. \ref{HeHm} corresponds to the NNI with

\begin{equation*}
V_{ij}=-\frac{\mu _{0}\mu ^{2}}{4\pi }\int d\mathbf{r}\int d\mathbf{\
	r^{\prime }}\left\vert \phi _{i}(\mathbf{r})\right\vert ^{2}\frac{1-3\cos
	^{2}\theta _{\mathbf{r}-\mathbf{r}^{\prime }}}{|\mathbf{r}-\mathbf{r}%
	^{\prime }|^{3}}\left\vert \phi _{j}(\mathbf{r^{\prime }})\right\vert ^{2}
\end{equation*}%
coming from the DDI (the contribution from the contact interaction is
proportional to the square of the exponentially small overlap and is
therefore neglected). Note that the DDI also generates interactions $V_{ij}$
beyond the nearest-neighbors, which decay as $\left\vert \mathbf{R}_{i}-\mathbf{R}_{j}\right\vert
^{-3} $. The corresponding terms are neglected in the Hamiltonian of Eq.\,5
because they are smaller and bring no new qualitative features to the
results of the present paper. Finally, the fourth term in Eq.\,5
describes the DAT with the amplitude $\Delta
J_{ij}=\Delta J_{ij,{\rm s}}+\Delta J_{ij,{\rm dd}}$ resulting from the contribution
from the contact interaction%
\begin{equation*}
\Delta J_{ij,{\rm s}}=-\frac{4\pi \hbar a_{s}}{m}\int d\mathbf{r}\left\vert \phi
_{i}(\mathbf{r})\right\vert ^{2}\phi _{i}^{\ast }(\mathbf{r})\phi _{j}(%
\mathbf{r})
\end{equation*}%
and from the dipole-dipole one%
\begin{equation*}
\Delta J_{ij,{\rm dd}}=-\frac{\mu _{0}\mu ^{2}}{4\pi }\int d\mathbf{r}\int d%
\mathbf{r^{\prime }}\left\vert \phi _{i}(\mathbf{r})\right\vert ^{2}\frac{%
	1-3\cos ^{2}\theta _{\mathbf{r}-\mathbf{r}^{\prime }}}{|\mathbf{r}-\mathbf{r}%
	^{\prime }|^{3}}\phi _{i}^{\ast }(\mathbf{r}^{\prime })\phi _{j}(\mathbf{r}%
^{\prime }).
\end{equation*}
It should be mentioned that in our experiments we also have a shallow confining potential $V_{\rm h}({\bf r})$. It can be taken into account by adding the term $\sum_i V_{{\rm h}, i}\, n_i$ with $V_{{\rm h}, i}  = \int d{\bf r} |\phi_i ({\bf r})|^2 V_{{\rm h}}({\bf r}) $ to the Hamiltonian of Eq.\,5.

The above expressions, together with numerically computed Wannier functions,
provide theoretical values for the parameters in the eBH model. During the
calculations, the singularity for $|\mathbf{r}-\mathbf{r}^{\prime
}|\rightarrow 0$ in the contributions from the DDI was
resolved by performing the integration over $\theta _{\mathbf{r}-\mathbf{r}%
	^{\prime }}$ before integrating over $|\mathbf{r}-\mathbf{r}^{\prime }|$.
For our experimental conditions, the contributions from the DDI is typically few times smaller than those from the
short-range interaction, but have strong dependence on the form of the Wannier
function $\phi _{i}(\mathbf{r})\,$\ and on the alignment of dipoles relative
to the lattice axes (see Fig.\,1, D to F).

\subsection*{SF-to-MI transition in the mean-field (MF) approximation}

In the MF approximation (see, e.g. \cite{Jaksch1998,Fisher1989,Sachdev2000}%
), the ground-state wavefunction of the system is written as a product
state over sites: 
\begin{equation*}
|\Psi _{G}\rangle =\bigotimes_{i}(\sum_{n=0}^{\infty }C_{i}^{(n)}|n\rangle
_{i}),
\end{equation*}%
where $|n\rangle _{i}$ denotes the Fock state with $n$ bosons on site $i$.
The coefficients $\{C_{i}^{(n)}\}$ are the variational parameters subjected
to the constraint $\sum_{n=0}^{\infty }\left\vert C_{i}^{(n)}\right\vert
^{2}=1$, which can be determined by minimizing the energy $E_{G}=\langle
\Psi _{G}|H|\Psi _{G}\rangle $. The SF phase is characterized by the local
order parameter $\left\langle b_{i}\right\rangle =\langle \Psi
_{G}|b_{i}|\Psi _{G}\rangle =\sum_{n=1}^{\infty }\sqrt{n}%
C_{i}^{(n-1)}C_{i}^{(n)}\neq 0$, which implies that $C_{i}^{(n)}$ are
non-zero for several adjacent values of $n$. In contrast, in the MI phase $%
C_{i}^{(n)}$ are non-zero for only one value of $n$. In a spatially
inhomogeneous system (e.g., in the presence of a trapping potential $V_{i}$),
this value is site-independent, and the system has typically a layered
structure in which the Mott states with different $n$ ($n=1$ and $2$ in our
case) are separated by the SF phase.

It should be mentioned that, even though the MF approximation is known to
overestimate the stability of the SF phase, here we are interested not in
the phase boundary of the SF-to-MI transitions itself, but in the relative
shift of this boundary when the dipolar polarization is changed from $\theta=0^\circ$ to $\theta=90^\circ$. In calculating this difference, the MF method is expected
to be much more reliable, as it is demonstrated by a good agreement between
theoretical and experimental results (Fig.\,4E).

\subsection*{Observability of the stripe phase}

The stripe phase is an example of exotic quantum phases induced by the NNI, which
is characterized by spontaneous translational symmetry breaking along one
direction. It can be accessed in a deep optical lattice half-filled with
atoms, when the NNI overwhelms the effects of single-particle tunneling and
temperature. In this case, the onsite interaction is much larger than all the
other parameters in the Hamiltonian, and prevents two atoms to be on the same
lattice site. We therefore can consider atoms as hardcore bosons, such that
the number of atoms on a lattice site can be only zero or one. We assume that the dipoles are polarized along
the $x$-direction resulting in attractive $V^{\rm att}$ (repulsive $V^{\rm rep}$) NNI
for the bonds $ij$ in the $x$($y$)-direction, with $-V^{\rm att}=2V^{\rm rep}=2V$. To
calculate the critical temperature for the stripe phase, we consider typical
experimental conditions with $s_{x}=s_{y}=20$ and ${\rm AR}=1$. In such a lattice, we
can ignore the DAT and the tunneling in the $z$-direction, and the
single-paticle tunneling amplitudes $J_{ij}$ do not depend on the direction of
the hopping, $J_{ij}=J=h\times20.5\mathrm{Hz}$. We obtain
$V=h\times34\mathrm{Hz}$ and the NNI value for the
bonds in the $z$-direction is $V/8=h\times4.25\mathrm{Hz}$. After neglecting this small coupling in the
$z$-direction, the Hamiltonian for each $xy$-plane can now be written as
\begin{equation}
H_{xy}=-J\sum_{\langle ij\rangle}(b_{i}^{\dag}b_{j}+h.c)+\sum_{i}%
(-2Vn_{i}n_{i+\hat{e}_{x}}+Vn_{i}n_{i+\hat{e}_{y}}-\mu
n_{i}),\label{eq:simHam}%
\end{equation}
where $b_{i}^{\dag}$ and $b_{i}$ are hard-core boson operators and $i+\hat{e}_{x}$ ($i+\hat{e}_{y}$) denotes the neighboring site of site $i$ in the $x$ ($y$) direction. We also add
the chemical potential $\mu$ which is chosen as $\mu=-V$ to satisfy the
condition of half-filling $\left\langle n_{i}\right\rangle =1/2$.  To
determine the critical temperature of the transition into the stripe phases,
we perform Quantum Monte Carlo calculations based on the worm algorithm for
the Hamiltonian of Eq.\,6, which is free from the negative sign
problem. For the above parameters, the calculated value for the critical
temperature is $T_{c}=1.4J\simeq1.5\mathrm{nK}$.

\bibliographystyle{apsrev}

\bibliography{Er_3DLattice_ref_w.o.URL,ultracold,theory}

\begin{thebibliography}{49}
\bibitem[{not({\natexlab{b}})}]{note6}
\bibinfo{note}{Note that the DAT have been recently observed for particles
  interacting only via the isotropic contact interaction
  \cite{Juergensen2014ood}.}
\bibitem[{not({\natexlab{c}})}]{note2}
\bibinfo{note}{For our lattice configuration the spherical point with
  $U_{\mathrm{dd}}=0$ is shifted to $\mathrm{AR}=1.05$ when including
  corrections to the harmonic approximation.}
\end{thebibliography}

\end{document}